\documentclass[3p,twocolumn]{elsarticle}

\usepackage{bm}
\usepackage{units}
\usepackage{amsmath}
\usepackage{textcomp}
\newcommand{\srm}[1]{_\mathrm{#1}}
\newcommand{\rmd}{\mathrm{d}}
\newcommand{\Arg}{\mathrm{Arg}}
\journal{Optics Communications}

\begin{document}

\begin{frontmatter}

\title{Measuring sub-Planck structural analogues in chronocyclic phase space}
\author{Dane R. Austin, Tobias Witting and Ian A. Walmsley}
\address{Clarendon Laboratory, University of Oxford, Parks Road, Oxford, OX1 3PU, UK }

\begin{abstract}
The phase space structure of certain quantum states reveals structure
on a scale that is small compared to the Planck area. Using an analog
between the wavefunction of a single photon and the electric field of
a classical ultrashort optical pulse we show that spectral shearing
interferometry enables measurement of such structures directly. 
Thereby extending the idea of Praxmeyer et al.  In particular, we use
multiple-shear spectral interferometry to fully characterize a pulse
consisting of two sub-pulses which are temporally and spectrally
disjoint, without a relative-phase ambiguity. This enables us to
compute the Wigner distribution of the pulse.  This spectrographic
representation of the pulse field features fringes that are tilted
with respect to both the time- and frequency axes, showing that in
general the shortest sub-Planck distances may not be in the directions
of the canonical (and easily experimentally accessible) directions.
Further, independent of this orientation, evidence of the sub-Planck
scale of the structure maybe extracted directly from the measured
signal.
\end{abstract}
\end{frontmatter}

\section{Classical and quantum wave fields}
\label{sec:intro}

The electromagnetic field is a proxy for the wavefunction of a single
photon \cite{Bialynicki-Birula-1994-Wave}. That is, a
spatio-temporally localized optical pulse field plays the role of a
probability density function (with suitable normalization) for the
measurement of a single quantum of light
\cite{Smith-2007-Photon}. Interference phenomena that are observed
with classical fields therefore bear a close relationship to quantum
interference phenomena for single particles, and this provides a route
to exploring quantum phase space structures in a simple way. It is not
perhaps surprising that there is a similarly close relationship
between the characterization of a classical ultrafast optical pulse
(i.e. the measurement of its electric field) and the quantum state
tomography of a single photon, and this analog may be used to apply
recently developed measurement techniques in ultrafast optics to the
study of quantum phase space.  Krzysztof Wodkiewicz exploited this
analog to effect in a number of directions, most recently in a study
of so-called sub-Planck phase-space structures discovered by Zurek.

The key elements of state tomography for localized photons are the
same as those required to fully characterize the electric field of a
short optical pulse. The central idea for this is related to the
Wodkiewicz-Eberly notion of the ``physical spectrum'', which is simply
the time-resolved intensity of a spectrally filtered pulse
\cite{Eberly-1977-time-dependent},
\begin{equation}
\label{phys_spec}
S(t, \omega) = \left| \int_{-\infty}^{t} \rmd t' K(t-t'; \omega) E(t') \right|^{2}
\end{equation}
with $E(t)$ the input pulsed electric field, and $K(t;\omega)$ the
time-stationary causal response function of the linear filter. Here,
$\omega$ is the passband frequency of the filter, which acts as a
spectrometer, and can be tuned. This formulation provides a basis for
developing methods for measuring $E(t)$, even using slow detectors
\cite{Walmsley-2009-Characterization}. In fact, the signal, considered
as a function of the two variables $t$ and $\omega$ is a sonogram of
the input pulse. That is, a positive-definite phase-space
representation of a field (as we consider only pulsed sources in this
paper, we shall assume the input field has compact support), obtained
by selecting a narrow spectral component of the pulse, and measuring
its time-dependent intensity after the filter. By adjusting the
passband frequency and measuring the temporal intensity at each
frequency, a time-frequency representation - the sonogram - is built
up.

More generally, the fast-responding detector that is required for
measuring Eberly-Wodkiewicz physical spectrum can be replaced by a
slow, time-integrating detector in front of which there is a
time-non-stationary linear filter, such as a fast shutter. This may be
modelled as a linear filter with response function $G(t;\tau)$, such
that the filter transmission is unity for a small time window near
$t=\tau$. The transmitted field after this "time gate" is
\begin{equation}
\label{sonogram}
E_{out}(t)= G(t;\tau) E_{in}(t).
\end{equation}
A more common approach to the characterization of ultrashort optical
pulses involves these two elements in reverse order. That is, a
temporal slice of the test pulse field is taken using a shutter, and
the transmitted spectrum measured using a spectrometer and a slow
detector. The resulting positive definite time-frequency
representation of the pulse is a spectrogram. In fact,
if the spectrometer passband is very narrow, the detected signal is
closely related to a Gabor transform
\begin{equation}
  S(\tau, \omega) = \left|\int_{-\infty}^{\infty} \rmd t
    G(t-\tau)E(t)e^{-i \omega t} \right|^{2}.
  \label{eq:Gabor}
\end{equation}
By adjusting the passband frequency and the delay $\tau$, over a
sufficient range, it can be shown that the detected signal is
tomographically complete. That is, it can be inverted to extract
$E(t)$. It is easy to see from the form of the signal that it has the
structure of a phase-space representation. Indeed, if $G(t)$ is a
Gaussian, the functional form is just that of the Husimi distribution
in quantum physics (or the Q-function in quantum optics) for a pure
quantum state. Therefore one might expect interference effects
parallel to those of interest in quantum systems to appear in the
spectrogram.

One approach to measuring the electric field based on spectrograms is
frequency-resolved optical gating (FROG)
\cite{Trebino-2000-Frequency-Resolved}, in which the ``gate function''
of (\ref{eq:Gabor}), $G(t)$, is implemented by means of a nonlinear
optical interaction. The test pulse field $E(t)$ is commonly mixed
with a time-delayed replica of itself in a material with a
$\chi^{(2)}$ nonlinear response. This generates the nonlinear
spectrogram
\begin{equation}
  S(\tau, \omega) = \left|  \int_{-\infty}^{\infty} \rmd t
    E(t-\tau) E(t) e^{-i \omega t} \right|^{2}.
  \label{FROG}
\end{equation}
Extracting the field from this nonlinear spectrogram requires an
iterative deconvolution algorithm \cite{Kane-1999-Recent}. However,
for the purposes of the measurements required in what follows this is
not necessary.

\begin{figure}[htb]
  \centering
  \includegraphics{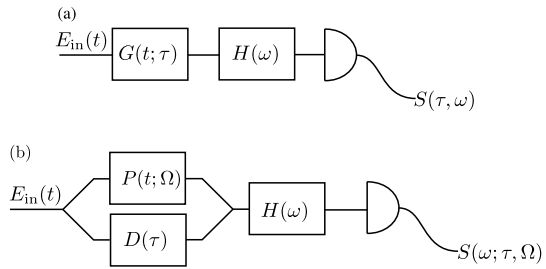}
  \caption{Linear filter representation of two pulse characterization
    schemes. (a) Gabor spectrogram, consisting of time-non-stationary
    gate $G(t;\tau)$ followed by spectral filter $H(\omega)$. (b)
    Spectral shearing interferometry, with a phase modulator producing a
    frequency shift $P(t;\Omega)$, delay line $D(\tau)$ and spectral
    filter $H(\omega)$.}
  \label{fig:filter_blocks}
\end{figure}

The Gabor spectrogram is a particular example of a general scheme for
pulse characterization that involves a sequence of two linear
filters. In that case, it is a time-non-stationary amplitude
filter (effectively a shutter) followed by a time-stationary amplitude
filter (a spectrometer), as depicted in
Fig.~\ref{fig:filter_blocks}(a). There are four possible minimal
sequences comprised of amplitude and phase-only filters.
These may be categorized both by the particular configuration of
filters and by the algorithm used to extract the field from the signal
\cite{Walmsley-1996-Characterization}.

A complementary approach is to use the filters in parallel. There are
similarly four possible configurations \cite{Iaconis98}
of apparatus in which the input pulses are split into two at a
beamsplitter, with each component traversing a filter, one of which is
time-stationary, and the other time non-stationary, after which the
filtered pulses are recombined and the spectrum of the interfering
pair registered.  A particular example of this category is spectral
shearing interferometry (SSI). In this case one of the filters
following the beamsplitter is a phase modulator. The role of this
non-stationary phase-only filter is to shift the frequency of the
pulse by an amount $\Omega$ --- the spectral shear. The parallel
filter has a time-stationary response, and causes a shift in the
arrival time of the second pulse replica --- it is a delay line. The
pulses are recombined on a second beamsplitter and the joint spectrum
measured. The linear filter representation of SSI is given in
Fig.~\ref{fig:filter_blocks}(b). The resulting spectral interference
pattern is
\begin{equation}
  S(\omega; \tau, \Omega) =
  | \tilde{E}(\omega)+ \tilde{E}(\omega - \Omega)e^{-i \omega \tau}|^{2}.
  \label{SPIDER}
\end{equation}
Again, when the modulator inducing the frequency shift is synthesized
using nonlinear optics, this approach is called spectral phase
interferometry for direct electric field reconstruction (SPIDER)
\cite{Iaconis-1998-Spectral}. The field reconstruction in this case is
linear and non-iterative. It is based on etracting the spectral phase
function of the field $\phi(\omega) = \Arg \tilde{E}(\omega)$.  A
simple Fourier transform of the signal $S(\omega; \tau, \Omega)$ with
respect to $\Omega$ enables the interference terms $
\tilde{E}^{\ast}(\omega) \tilde{E}(\omega - \Omega)e^{-i \omega \tau}$
to be separated, and the argument of this term, with a reference phase
subtracted, is simply $\Gamma(\omega) = \phi(\omega) -
\phi(\omega-\Omega)$. This sampling of the phase function is
sufficient to enable the pulse field to be accurately estimated on the
temporal domain $[ - \pi/ \Omega, \pi/\Omega ]$.

Using both FROG and SPIDER is it is possible to construct a Wigner
phase space representation of the pulse directly. It is this function
that is commonly used in quantum mechanics to study nonclassical
properties of quantum states. Of particular interest are the types of
``sub-Planck'' structures discussed by Zurek
\cite{Zurek-2001-Sub-Planck}. He showed that two states with
significant overlap in phase space may be in fact orthogonal, even if
they are separated in either dimension by a degree that is
significantly smaller than the scale implied by Planck's constant.

Using the measurement methods discussed above, it is also possible to
extract directly a signal component that illustrates the important
structure of the field. This is a considerably more efficient approach
than a full characterization of the field, or, equivalently for single
photons, full state tomography. In effect, these components are
``witnesses'' to the sub-Planck structure of the field. In the next
section we introduce sub-Planck structure and describe the
measurements necessary for implementing witnesses.

\section{Sub-Planck structure}
\label{sec:sub-Planck}

One measure of the sensitivity of a quantum state to perturbations is
the size of the smallest displacement $\bm{\delta}\srm{q} = (\delta_x,
\delta_p)$ in phase space required to make the state orthogonal to the
original; that is, for the inner product
\begin{equation}
  \langle \psi \mid \psi' \rangle =
  \int_{-\infty}^{\infty}\phi^{\ast}(x)e^{i \delta_p x / \hbar}
  \phi(x + \delta_x) \rmd x
  \label{eq:quan_overlap}
\end{equation}
between state $\mid \psi \rangle$ and its displaced replica $\mid
\psi' \rangle$ to be zero. Zurek showed that this minimum displacement
along a given direction in phase space is inversely proportional to
the spread of the state in the perpendicular direction. For example,
the lowest-order estimate of the ``zero-overlap'' momentum shift
$\delta_p$ is
\begin{equation}
  \delta_p \approx \frac{\hbar}{\sqrt{\langle x^2 \rangle - 
      \langle x \rangle^2}},
  \label{eq:rms_spread}
\end{equation}
i.e. inversely proportional to the root-mean square spread of the
state. The size of the zero-overlap shift can be significantly smaller
than the Planck scale $\sqrt{\hbar}$. This is connected to the
existence of phase space structures of dimension given by
(\ref{eq:rms_spread}) which appear in the Wigner representation of the
state,
\begin{equation}
W(x,p)=\frac{1}{2\pi\hbar}\int_{-\infty}^{\infty} e^{i p y /\hbar} 
\psi^{\ast}(x-\frac{y}{2})\psi(x-\frac{y}{2}) \rmd y.
\end{equation}
By means of arbitrary rotations of phase space, this argument can be
applied along any direction in phase space: the smallest displacement
leading to orthogonality and the sub-Planck structure both have a size
inversely proportional to the spread of the wave function along the
perpendicular direction.

\begin{figure}[htb]
  \centering
  \includegraphics{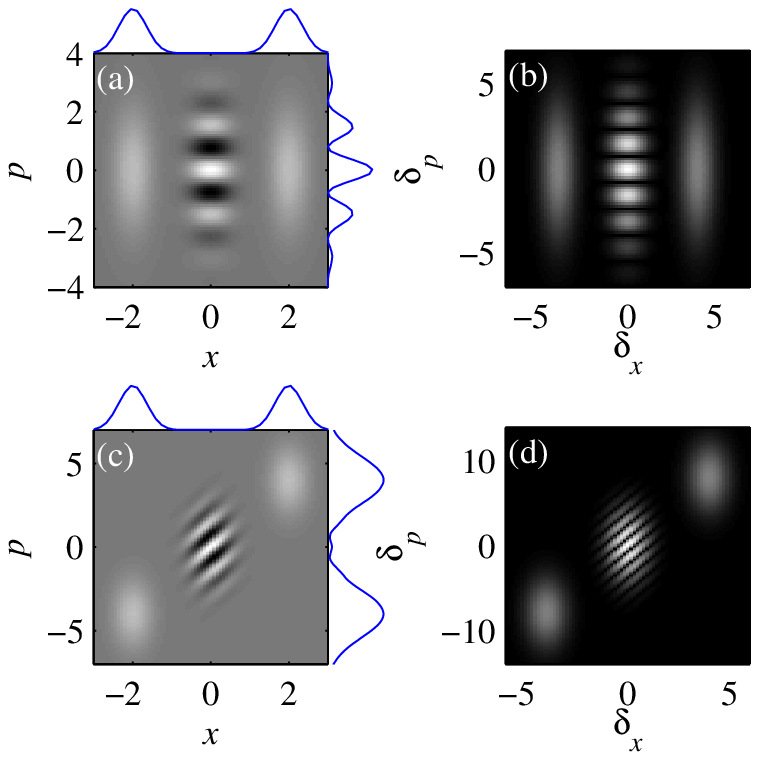}
  \caption{Wigner and overlap functions of quantum states consisting
    of two separate Gaussians. (a) Wigner function, Gaussian states
    separate in position. The blue lines show the marginals. (b)
    Overlap function corresponding to (a). (c) Wigner function with
    the Gaussian states separate in position and momentum. (d) Overlap
    function corresponding to (b).}
  \label{fig:Wigner_overlap}
\end{figure}

Some examples of Wigner distributions are given in
Fig.~\ref{fig:Wigner_overlap}. We use atomic units, with $\hbar =
1$. Fig.~\ref{fig:Wigner_overlap}(a) shows a wavefunction consisting
of a superposition of two Gaussian states separated position by $x_0 =
4$. Sub-Planck structure is evident, in the form of interference
fringes along the momentum axis.
The magnitude of the overlap as a function of $(\delta_x,\delta_p)$ is
shown in Fig.~\ref{fig:Wigner_overlap}(b). The overlap also contains
fringes, modulated proportional to $1 + \cos(\delta_p r_0)$. The
smallest shift which produces a zero-overlap, and hence orthogonality,
occurs for $\delta_p = \pm \pi/r_0$. For the clarity of
Fig.~\ref{fig:Wigner_overlap}, we did not choose $r_0$ to be
particuarly large, but it is clear that as $r_0$ increases both
the fringe period and area can decrease well below their respective
Planck length and area scales.

The sub-Planck structure of the the analogous classical optical field,
with time and frequency substituted for position and momentum, has
been studied by W\'odkiewicz and coworkers
\cite{Praxmeyer-2007-Time-Frequency}. The Wigner distribution of a
double pulse, consisting of two identical sub-pulses with time delay
$t_0$, contains spectral interference fringes of period $2\pi/t_0$
To directly observe the scalar product, W\'odkiewicz and others
pointed out that the spectrum of the second harmonic of an ultrashort
pulse
\begin{equation}
  I_2(\omega) = \left| \int_{-\infty}^{\infty} \rmd t E^2(t) e^{i \omega t} \right|^2
\end{equation}
is equivalent to the scalar product of the pulse with a frequency shifted replica
\begin{equation}
  \left| \langle E(t) |  E(t)e^{i \delta_\omega t} \rangle \right |^2 
  = \left| \int_{-\infty}^{\infty} E^{\ast}(t) E(t) e^{i \delta_\omega t} \right|^2
  \label{eq:FROGoverlap}
\end{equation}
when the electric field has flat temporal phase so that the presence
of the complex conjugate in (\ref{eq:FROGoverlap}) makes no
difference. Using Frequency-Resolved Optical Gating (FROG) they
directly observed the scalar product for various frequency
displacements.

This arrangement enables the direct observation of the overlap between
a pulse and its frequency-shifted replica, with the additional proviso
that the pulse has a linear temporal phase. In many circumstances it
is desireable to measure the overlap with arbitrary linear
combinations of time- and frequency shifts, and for arbitrary
pulses. Fig.\ref{fig:Wigner_overlap}(c) shows a superposition of
Gaussian states which are displaced in both position and momentum by
$\mathbf{r}_0 = (x_0,p_0) = (4,8)$. Notably, although the individual
states do not overlap in either domain, interference fringes appear in
the Wigner distribution aligned parallel to the separation vector.
The amplitude of the overlap is shown in Fig.~\ref{fig:Wigner_overlap}(c). 
In this domain, the central fringe modulation is proportional to $1 +
\cos(\bm{\delta}\srm{q} \cdot \mathbf{k}_0)$, where the wavenumber
$\mathbf{k}_0 = (p_0, -x_0)$ is perpendicular to the separation
$\mathbf{r}_0$. The smallest shift which produces a zero is $\pm \pi
\mathbf{k}_0/|\mathbf{k}_0|^2 = \pm \pi (-p_0,x_0) / (x_0^2 + p_0^2)$,
again perpendicular to $\mathbf{r}_0$.

We shall now show that the scalar product between a pulse and a time-
and frequency-shifted replica is given by the same raw data which is
used in SSI. As described in Section~\ref{sec:intro} above, in SSI the
interferometric term
\begin{equation}
  D(\omega, \Omega) = \tilde{E}^{\ast}(\omega)\tilde{E}(\omega-\Omega)
\end{equation}
is obtained via Fourier filtering of the interferogram and removal of
the carrier phase. For normal pulse reconstruction, the phase of
$D(\omega, \Omega)$ is used to reconstruct the spectral phase of the
unknown pulse. However, here we note that by setting $\Omega =
\delta_\omega$, the Fourier transform of $D(\omega, \delta_\omega)$
along $\omega$ is given by
\begin{eqnarray}
  \hspace{-3mm}\tilde{D}(\delta_t,\delta_\omega)
  & = & \int_{-\infty}^{\infty} E(t) E^\ast(t-\delta_t)e^{-i \delta_\omega t} \rmd t \\
  & = & \langle E(t-\delta_t)e^{i \delta_\omega t}|E(t)\rangle
  \label{eq:SSIoverlap}
\end{eqnarray}
which is exactly the overlap between the field and replica shifted
$\delta_\omega$ in frequency and $\delta_t$ in time. Therefore, the
Fourier transform of the filtered sideband provides the overlap
directly without any intermediate processing. Zeros of
$\tilde{D}(\delta_t,\delta_\omega)$ around the origin are witnesses to
the sub-Planck structure of the quantum state analogous to the unknown
pulse.

In the next section, we describe an experimental setup for observing
witnesses to the sub-Planck structure in an ultrashort pulse analagous
to the position- and momentum-separated quantum state in
Fig.~\ref{fig:Wigner_overlap}(c).

\section{Experiment}
\begin{figure}
  \centering
  \includegraphics{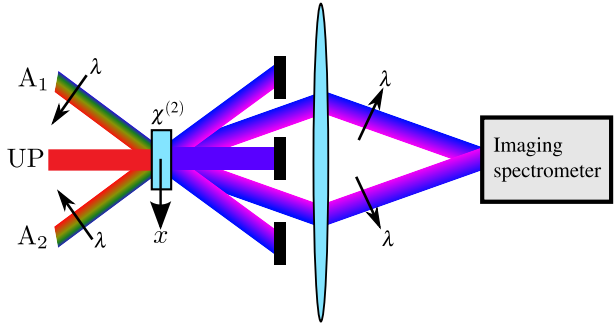}
  \caption{SEA-CAR-SPIDER concept; the unknown pulse (UP) undergoes
    sum-frequency mixing with two oppositely spatially chirped
    ancillae (A$_1$, A$_2$) in a $\chi^{(2)}$ crystal. The upconverted
    beams are re-imaged onto a two-dimensional imaging spectrometer.}
  \label{fig:seacarsetup}
\end{figure}

There are many implementations of SSI that can be used to measure
$D(\omega, \delta_\omega)$ for single frequency shifts. For
convenient, single-shot measurement of a continuous range of frequency
shifts, we used SEA-CAR-SPIDER \cite{Witting-2009-Ultrashort}, a
recent twist on SSI that permits the simultaneous acquisition of
multiple shears and a precise and robust calibration procedure. The
principle of SEA-CAR-SPIDER is shown in Fig.~\ref{fig:seacarsetup}.
Two spatially chirped fields, known as the ancillae, are prepared and
sum-frequency mixed with the unknown pulse in a $\chi^{2}$
crystal. The the ancilla beams are oppositely spatially chirped, so that in the
crystal their local frequencies at transverse position $x$ (indicated
in Fig.~\ref{fig:seacarsetup}) are $\omega\srm{up} \pm \alpha
x$. Here, $\omega\srm{up}$ is their common upconversion frequency at
$x=0$ and $\alpha$ is the amount of spatial chirp. Two upconverted
replicas of the unknown pulse are produced, with a relative spectral
shear of $\delta_\omega = 2\alpha x$. These are re-imaged onto an
imaging spectrometer, their converging angle producing a spatial
interference pattern
\begin{equation} 
  S(\omega + \omega\srm{up},x) = \left| E(\omega + \alpha x) e^{i k_c x}
    + E(\omega - \alpha x) \right|^2
  \label{eq:SEACARsig}
\end{equation}
where $k\srm{c}$ is the wavenumber of the interference fringes,
determined by the beam convergence angle. To process this
interferogram, either for reconstruction of the phase of the unknown
pulse or for determination of the overlap (\ref{eq:SSIoverlap}), we
apply the two-dimensional Fourier transform to the $(t,k_x)$
domain. Here, the transforms of the spectral intensities $|E(\omega
\pm \alpha x)|^2$ appear at the origin, whilst the interference term
between the replicas manifests as two sidebands located at $(0,\pm
k\srm{c})$. One of these is isolated and the inverse transform
applied. Removing the spatial carrier $k\srm{c}x$ and using the linear
relation between $x$ and the shear, we obtain
$D(\omega,\delta_\omega)$.

The optical pulse analogue to the quantum state of
Fig.~\ref{fig:Wigner_overlap} is a bichromatic double pulse --- i.e. two
sub-pulses with both a temporal and spectral separation. We
synthesized these using a \mbox{4-f} pulse shaper with a $\approx
\unit[1.1]{mm}$ thick glass slide in one half of the Fourier plane, so
that frequencies above $\unit[2.287]{rad~fs^{-1}}$ were delayed. We
also placed a thin opaque block at the edge of the glass slide,
attenuating frequencies from $\unit[2.284 --
2.291]{rad~fs^{-1}}$. Through this procedure, we created a bichromatic
double pulse with temporal separation $t_0 = \unit[1820]{fs}$ and
spectral separation $\omega_0 = \unit[18]{mrad~fs^{-1}}$, which we
then characterized using the SEA-CAR-SPIDER apparatus.

\begin{figure}[htb]
  \centering
  \includegraphics{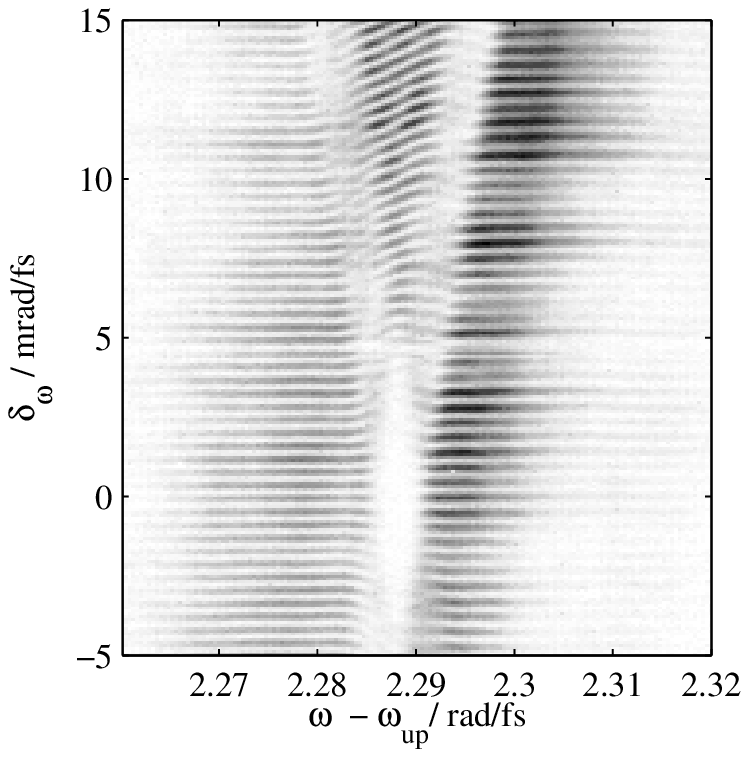} 
  \caption{Raw SEA-CAR-SPIDER data for the bichromatic double pulse.}
  \label{fig:raw}
\end{figure}

The raw data trace is shown in Fig.~\ref{fig:raw}. The fringe pattern can be
interpreted as follows: the phase of each fringe is modulated by
$\phi(\omega + \delta_\omega) - \phi(\omega - \delta_\omega)$. To
either side of the spectral gap at $\unit[2.287]{rad~fs^{-1}}$, the
two frequencies $\omega \pm \delta_\omega$ both reside in the same
sub-pulse. Each sub-pulse is transform limited, and hence the phase
difference between the two frequencies is a constant. The fringes are
therefore unmodulated away from the spectral gap. Closer to the
spectral gap, and for larger shears, $\omega \pm \delta_\omega$ reside
in different sub-pulses. Their different group delay produces a phase
difference which varies linearly with frequency. The fringes are
therefore tilted in this region.

\begin{figure}[htb]
  \centering
  \includegraphics{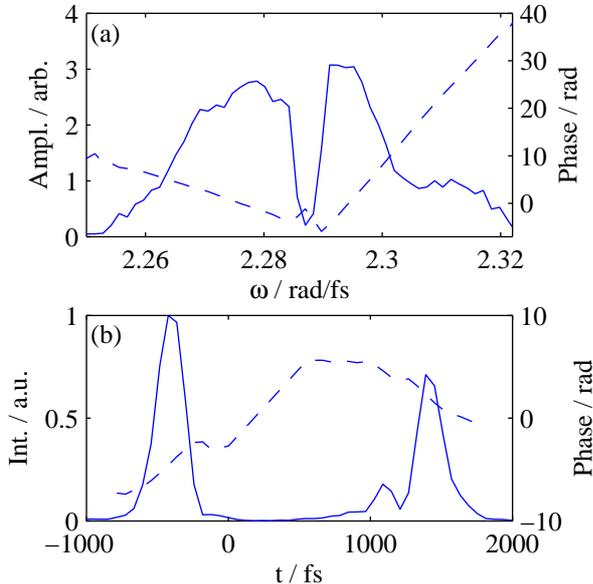}
  \caption{(a) Spectral amplitude (solid, left axis) and phase
    (dashed, right axis) of the bichromatic double pulse.  (b)
    Corresponding temporal intensity (solid, left axis) and phase
    (dashed, right axis).}
  \label{fig:spectrum}
\end{figure}

\begin{figure}[htb]
  \centering
  \includegraphics{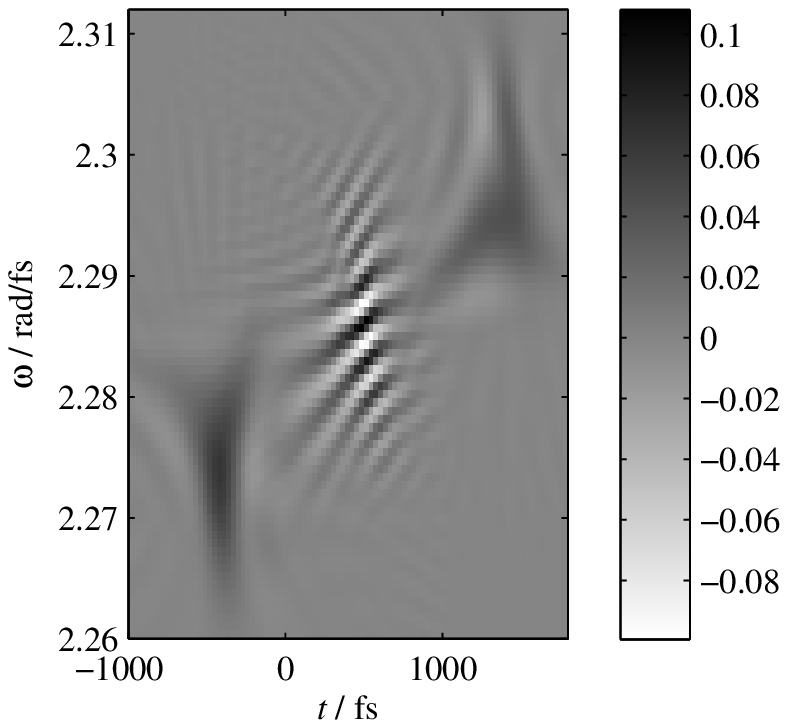}
  \caption{Wigner distribution of the bichromatic double pulse.}
  \label{fig:Wigner}
\end{figure}

Although our aim was to directly measure the overlap
(\ref{eq:SSIoverlap}), we also reconstructed the pulse to verify our
synthesis and to numerically calculate its Wigner function. Because
the pulse contained significant gaps in its spectrum, it was necessary
to use the multiple shear reconstruction algorithm,
\cite{Austin-2009-High}, a recently developed generalization of
standard SSI concatenation procedure \cite{Iaconis-1998-Spectral}
which enables the use of a large spectral shear to bridge spectral
gaps. The measured spectrum and temporal profile of the bichromatic
double pulse is shown in Fig.~\ref{fig:spectrum}.
We calculated the Wigner distribution of the measured pulse, shown in
Fig.~\ref{fig:Wigner}. The two peaks at
$(t,\omega)=(\unit[-423]{fs},\unit[2.275]{rad/fs})$ and
$(\unit[1387]{fs}, \unit[2.293]{rad/fs})$ represent the sub-pulses,
whilst the interference fringes between them are aligned parallel to
their displacement $(t_0,\omega_0)$ in phase space. As expected,
Fig.~\ref{fig:Wigner} bears a close qualitative similarity to
Fig.~\ref{fig:Wigner_overlap}(c).

\begin{figure}[htb]
  \centering
  \includegraphics{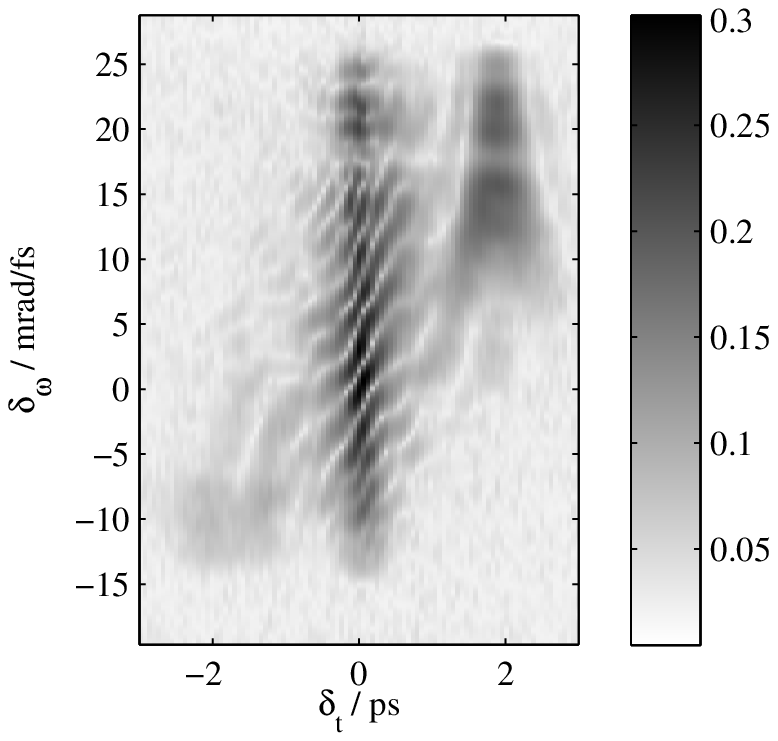}
  \caption{Phase space autocorrelation of the bichromatic double pulse.}
  \label{fig:psauto}
\end{figure}

We calculated the overlap of the pulse and its phase-space-displaced
replica $\tilde{D}(\delta_t, \delta_\omega)$ by Fourier transforming
$D(\omega,\delta_\omega)$. The result is shown in
Fig.~\ref{fig:psauto}, and as expected bears a close similarity to
Fig.~\ref{fig:Wigner_overlap}(d), with fringe modulation proportional
to $1 + \cos \left[ (\delta_\omega,\delta_t) \cdot (\omega_0,-t_0)
\right]$. Zero overlap occurs whenever $(\delta_\omega,\delta_t) \cdot
(\omega_0,-t_0) = \pm \pi$.

\section{Summary and conclusions}

The analogy between the electric field of an ultrashort pulse and the
wavefunction of a single photon can be exploited so that classical
wave interference in optics mimics quantum interference, enabling
observation of phase-space structures that have important implications
for dissipative quantum dynamics.

Sub-Planck structure is just such a quantum interference effect which
determines the sensitivity of a quantum state to perturbations,
considered as displacements in phase space. We have demonstrated that
spectral shearing interferometry directly measures the overlap between
an ultrashort pulse and a replica that has displaced in phase space in
a range of directions, thereby providing a direct test of the Zurek
hypothesis.

An important question is whether it might be possible to implement the
same approach for a quantum system consisting of massive particles? In
fact, it is possible to devise a shearing interferometer for the
center-of-mass wavefunction of an atom. \cite{Walmsley99} A momentum
shift of the transverse degree of freedom can be accomplished by
passing the atom through a standing-wave light field in the Lamb-Dicke
limit. Propagation in free space leads to the detection of an atom
interferogram with similar properties to those of the SPIDER method
discussed here.

Such methods may enable a deeper exploration of uniquely quantum
characteristics, those which distinguish purely quantum dynamics from
that possible in classical systems, to which question Kryszstof
Wodiewicz devoted a great deal of his work.

\bibliographystyle{elsarticle-num}

\begin{thebibliography}{10}
\expandafter\ifx\csname url\endcsname\relax
  \def\url#1{\texttt{#1}}\fi
\expandafter\ifx\csname urlprefix\endcsname\relax\def\urlprefix{URL }\fi
\expandafter\ifx\csname href\endcsname\relax
  \def\href#1#2{#2} \def\path#1{#1}\fi

\bibitem{Bialynicki-Birula-1994-Wave}
I.~Bialynicki-Birula, On the wave function of the photon, Acta Phys. Pol., A 86
  (1994) 97--116.

\bibitem{Smith-2007-Photon}
B.~J. Smith, M.~G. Raymer, \href{http://stacks.iop.org/1367-2630/9/414}{Photon
  wave functions, wave-packet quantization of light, and coherence theory}, New
  J. Phys. 9~(11) (2007) 414.
\newline\urlprefix\url{http://stacks.iop.org/1367-2630/9/414}

\bibitem{Eberly-1977-time-dependent}
J.~H. Eberly, K.~W\'{o}dkiewicz,
  \href{http://www.opticsinfobase.org/abstract.cfm?URI=josa-67-9-1252}{The
  time-dependent physical spectrum of light}, J. Opt. Soc. Am. 67~(9) (1977)
  1252--1261.
\newline\urlprefix\url{http://www.opticsinfobase.org/abstract.cfm?URI=josa-67-%
9-1252}

\bibitem{Walmsley-2009-Characterization}
I.~A. Walmsley, C.~Dorrer,
  \href{http://aop.osa.org/abstract.cfm?URI=aop-1-2-308}{Characterization of
  ultrashort electromagnetic pulses}, Adv. Opt. Photon. 1~(2) (2009) 308--437.
\newline\urlprefix\url{http://aop.osa.org/abstract.cfm?URI=aop-1-2-308}

\bibitem{Trebino-2000-Frequency-Resolved}
R.~Trebino, Frequency-Resolved Optical Gating: The Measurement of Ultrashort
  Laser Pulses, Kluwer Academic Publishers, 2000.

\bibitem{Kane-1999-Recent}
D.~J. Kane, Recent progress toward real-time measurement of ultrashort laser
  pulses, IEEE J. Quantum Electron. 35~(4) (1999) 421, 0018-9197.

\bibitem{Walmsley-1996-Characterization}
I.~Walmsley, V.~Wong, {Characterization of the electric field of ultrashort
  optical pulses}, J. Opt. Soc. Am. B 13~(2453) (1996) 233.

\bibitem{Iaconis98}
C.~Iaconis, V.~Wong, I.~A. Walmsley, Direct interferometric techniques for
  characterizing ultrashort optical pulses, IEEE JSTQE 4 (1998) 1.

\bibitem{Iaconis-1998-Spectral}
C.~Iaconis, I.~A. Walmsley, Spectral phase interferometry for direct
  electric-field reconstruction of ultrashort optical pulses, Opt. Lett.
  23~(10) (1998) 792--794.

\bibitem{Zurek-2001-Sub-Planck}
W.~H. Zurek, Sub-planck structure in phase space and its relevance for quantum
  decoherence, Nature 412~(6848) (2001) 712--717.

\bibitem{Praxmeyer-2007-Time-Frequency}
L.~Praxmeyer, P.~Wasylczyk, C.~Radzewicz, K.~W\'{o}dkiewicz,
  \href{http://link.aps.org/abstract/PRL/v98/e063901}{Time-frequency domain
  analogues of phase space sub-planck structures}, Phys. Rev. Lett. 98~(6)
  (2007) 063901.
\newblock \href {http://dx.doi.org/10.1103/PhysRevLett.98.063901}
  {\path{doi:10.1103/PhysRevLett.98.063901}}.
\newline\urlprefix\url{http://link.aps.org/abstract/PRL/v98/e063901}

\bibitem{Witting-2009-Ultrashort}
T.~Witting, D.~R. Austin, I.~A. Walmsley,
  \href{http://arxiv.org/abs/0908.1245v1}{{Ultrashort pulse characterization by
  spectral shearing interferometry with spatially chirped ancillae}}, Arxiv
  preprint arXiv:0908.1245v1 [physics.optics].
\newline\urlprefix\url{http://arxiv.org/abs/0908.1245v1}

\bibitem{Austin-2009-High}
D.~R. Austin, T.~Witting, I.~A. Walmsley,
  \href{http://josab.osa.org/abstract.cfm?URI=josab-26-9-1818}{High precision
  self-referenced phase retrieval of complex pulses with multiple-shearing
  spectral interferometry}, J. Opt. Soc. Am. B 26~(9) (2009) 1818--1830.
\newline\urlprefix\url{http://josab.osa.org/abstract.cfm?URI=josab-26-9-1818}

\bibitem{Walmsley99}
I.~A. Walmsley, N.~P. Bigelow, Measuring the quantum state of cold atoms using
  momentum-shearing interferometry, Phys. Rev. A 57 (1998) R713.

\end{thebibliography}

\end{document}